\documentclass[11pt,
showpacs,
preprintnumbers,amsmath,amssymb,
aps,prd,nofootinbib,eqsecnum,a4paper]{revtex4}
\usepackage{epsfig}
\usepackage{graphicx,epsf}
\usepackage{color}
\usepackage{bm}
\usepackage{psfrag}
\usepackage[symbol]{footmisc}
\usepackage{tensor}
\usepackage{hyperref}
\def\beq{\begin{equation}}
\def\eeq{\end{equation}}
\def\bea{\begin{eqnarray}}
\def\eea{\end{eqnarray}}
\def\bi{\begin{itemize}}
\def\ei{\end{itemize}}
\def\cs2{c_{\rm{s}}^2}
\def \beg {\begin{enumerate}}
\def \en {\end{enumerate}}

\def\M0{{\cal M}_0}

\DeclareMathOperator{\sech}{sech}
\DeclareMathOperator{\arcsech}{arcsech}
\begin{document}

\title{Constraining $\alpha$-attractor models from reheating}

\author{Gabriel Germ\'an\footnote[1]{\href{mailto:gabriel@icf.unam.mx}{gabriel@icf.unam.mx}}}
\affiliation{
$^*$Instituto de Ciencias F\'{i}sicas, Universidad Nacional Aut\'{o}noma de M\'{e}xico,\\Av. Universidad S/N. Cuernavaca, Morelos, 62210, M\'{e}xico}
%

\begin{abstract}
We eliminate the parameters originally present in models of inflation of the $\alpha$-attractor type in favor of the scalar spectral index $n_s$ and the tensor-to-scalar ratio $r$. We then write expressions for the number of $e$-folds during reheating $ N_ {re} $. By imposing reasonable conditions on $N_{re}$ we can restrict $n_s$ and $r$ and in turn, we use these constraints in order to find bounds for cosmological quantities of interest such as the number of $e$-folds during inflation and the radiation dominated eras, as well as for the reheating temperature and the running index. The minimum condition that $N_ {re}$ must satisfy is $N_ {re}\geq 0$ which we use to constrain the cosmological quantities mentioned above. In particular, we find that the tensor-to-scalar ratio $r$ (and as a consequence the energy scale of inflation) is bounded from below. We provide figures illustrating the behavior of these quantities as functions of $r$ for several values of $n_s$ and tables containing the bounds so obtained.
  
\end{abstract}


\maketitle

\section {\bf Introduction}\label{Intro}
Inflationary models offer the possibility of better understanding the evolution of the early universe resulting in a plausible explanation of the present (for reviews see e.g., \cite{Linde:1984ir}-\cite{Martin:2018ycu}). A plethora of models have been proposed  \cite{Martin:2013tda} and there are several with very interesting characteristics that make them viable models. The post-inflationary stage of reheating has, however, been studied with less success, although there are several very interesting contributions to the subject (for reviews on reheating see e.g., \cite{Bassett:2005xm},  \cite{Allahverdi:2010xz}, \cite{Amin:2014eta}). 

In this article we would like to explore a fruitful connection between the reheating and the inflationary stages. In particular, we propose a procedure that allows to find bounds for cosmological quantities of interest by imposing restrictions on the number of $e$-folds during the reheating stage. For this, we write $ N_ {re} $ (the number of $e$-folds during reheating) in terms of the scalar spectral index $n_s$ and the tensor-to-scalar ratio $r$ eliminating the parameters originally present in the potential that defines the model under consideration. We write a general expression for $ N_ {re} $ whose evaluation requires the specification of an inflationary model both at horizon crossing and at the end of inflation. As illustrative examples we study a class of $\alpha$-attractor models \cite{German:2021rin} such that their potential is quadratic in the inflaton around the minimum of the potential where reheating takes place (for related work see e.g., \cite{Eshaghi:2016kne}, \cite{Ueno:2016dim}, \cite{DiMarco:2017zek}, \cite{Ellis:2021kad}). In this case, the equation of state parameter (EoS) during reheating is $\omega_{re}=0$.   By imposing the condition $N_ {re}\geq 0$ we obtain a restricted range for $n_s$ and $r$ and with this range we can then constrain cosmological quantities such as the number of $e$-folds during inflation $N_{ke}$, reheating $N_{re}$ and radiation $N_{rd}$ as well as the reheat temperature $T_{re}$ and the running index $n_{sk}$. It is found in particular that the tensor-to-scalar ratio $r$ (and thus, the energy scale of inflation) is bounded from below.

The organization of the article is as follows: in Section \ref{GEN} we discuss the strategy obtaining general expressions which will  be used later in the study of particular models. The most important equations in this section are given by Eqs.~\eqref{nsk} to \eqref{NREfinal} from which the bounds for cosmological quantities follow. In Section \ref{MOD} we study three examples from a class of models of the $\alpha$-attractor type \cite{German:2021rin} and the results are presented in Figs.~\ref{Nre124} and \ref{nsk124} and the tables \ref{bounds124} and \ref{0.9649}. In Section \ref{CONSISTENCY} we compare the results obtained with previous related work and show the consistency between the different procedures. Finally we conclude in Section \ref{CON}.
\section {\bf General strategy }\label{GEN} 

In this section we discuss the general  strategy that can be used in the study of inflationary models of interest. The key point is to eliminate the parameters, present in the potential defining the model of inflation, in terms of the observables $n_s$ and $r$. First we need to establish the equations relating the observables with the parameters of the model. By writing cosmological quantities in terms of $n_s$ and $r$ we can constrain them with the observational and/or theoretical bounds for $n_s$ and $r$. As $n_s$ and $r$ get measured with higher precision our quantities written in terms of them will also be bounded even further. 

Observables are given in the slow-roll (SR) approximation as follows (see e.g.  \cite{Lyth:1998xn}, \cite{Liddle:1994dx})
\begin{eqnarray}
n_{t} &=&-2\epsilon = -\frac{r}{8} , \label{Int} \\
n_{s} &=&1+2\eta -6\epsilon ,  \label{Ins} \\
n_{tk} &=&4\epsilon\left( \eta -2\epsilon\right)=\frac{r}{8}\left(n_s-1+\frac{3}{8}r\right), \label{Intk} \\
n_{sk} &=&16\epsilon \eta -24\epsilon ^{2}-2\xi_2, \label{Insk} \\
A_s(k) &=&\frac{1}{24\pi ^{2}\epsilon} \frac{V_k}{M_{pl}^{4}}. \label{IA} 
\end{eqnarray}
Equations \eqref{Int} and \eqref{Intk} are actually consistency conditions among the observables. Here the running of the scalar index $n_{s}$ is denoted by $n_{sk} \equiv \frac{d n_{s}}{d \ln k}$ and the running of the tensor spectral index $n_{t}$ by $n_{tk}\equiv \frac{d n_{t}}{d \ln k}$,  in a self-explanatory notation. In the literature $n_{sk}$ is usually denoted by $\alpha$ but here we prefer to use this more symmetrical notation between scalar and tensorial quantities. The amplitude of scalar density perturbations at wave number $k$ is $A_s(k)$. All quantities in Eqs.~\eqref{Int} to  \eqref{IA} are evaluated at
the pivot scale $k=k_p.$\footnote{The subindex $k$ or $k_p$ above denotes the value of the inflaton when scales the size of the pivot scale leave the horizon.} 
The SR parameters appearing above are defined by
\begin{equation}
\epsilon \equiv \frac{M_{pl}^{2}}{2}\left( \frac{V^{\prime }}{V }\right) ^{2},\quad
\eta \equiv M_{pl}^{2}\frac{V^{\prime \prime }}{V}, \quad
\xi_2 \equiv M_{pl}^{4}\frac{V^{\prime }V^{\prime \prime \prime }}{V^{2}}.
\label{SR}
\end{equation}
Also,  $M_{pl}\equiv 1/\sqrt{8\pi G}$ is the reduced Planck mass $M_{pl}=2.44\times 10^{18} \,\mathrm{GeV}$, primes on $V$ denote derivatives with respect to the inflaton field $\phi$.
From Eqs.~\eqref{Int} and \eqref{Ins} we find 
\begin{equation}
\eta =\frac{1}{16}\left( 3 r-8 \delta_{n_s}\right),
\label{Ieta}
\end{equation}
where $\delta_{n_s}$ is defined as $\delta_{n_s}\equiv 1-n_s$.  For the range of values for the scalar spectral index $0.9607<n_s<0.9691$ \cite{Akrami:2018odb} and the tensor-to-scalar ratio $r<0.036$ \cite{BICEPKeck:2021gln}, $\eta$ is bounded as $-0.01965<\eta<-0.0087$ thus, at horizon crossing the potential should be concave downwards. Using Eqs.~\eqref{Ieta} and \eqref{Int}, the expression for the running of the scalar index given by Eq.~\eqref{Insk} can be written as
\begin{equation}
n_{sk} =\frac{3}{32}r^2 - \frac{1}{2} \delta_{n_s} r -  \frac{1}{4} r \frac{V^{\prime \prime \prime }}{V^{\prime}}M_{pl}^2 \;.
\label{nsk}
\end{equation}
For models involving one or two parameters it is sometimes possible to eliminate them in terms of $n_s$ and $r$ by solving Eqs.~\eqref{Int}, \eqref{Ins} and \eqref{IA}. This has been done in \cite{German:2020eyq} where $V^{\prime \prime \prime}/V^{\prime}$ (thus, the running) has been written purely in terms of $n_s$ and $r$ for some other models of interest. Here we would like to apply a similar strategy to a class of $\alpha$-attractor models recently proposed by the author \cite{German:2021rin} to investigate possible restrictions coming from conditions imposed on the number of $e$-folds during reheating $N_{re}$. Thus, it is necessary first to write $N_{re}$ in terms of $n_s$ and $r$.   We will see that imposing the constraint $N_{re}\geq c$, where $c\geq 0$ is some well motivated bound for $N_{re}$, both $n_s$ and $r$ are further restricted and consequently new constraints are found for quantities of interest such as the number of $e$-folds during inflation $N_{ke}$, radiation $N_{rd}$, the running $n_{sk}$ and the reheat temperature $T_{re}$. 

Building on previous work \cite{Liddle:2003as}, \cite{Dodelson:2003vq}, \cite{Liddle:1994dx}, it is possible to find an expression for the number of $e$-folds during reheating  \cite{Dai:2014jja}, \cite{Munoz:2014eqa}  as follows (see also e.g., section 3 of \cite{German:2020iwg})
\beq
\label{NRE}
N_{re}= \frac{4}{1-3\, \omega_{re}}\left(-N_{ke}-\frac{1}{3} \ln[\frac{11 g_{s,re}}{43}]-\frac{1}{4} \ln[\frac{30}{\pi^2 g_{re} } ] -\ln[\frac{\rho^{1/4}_e k}{H_k\, a_0 T_0} ]\right),
\eeq
where $\omega_{re}$ is the EoS at the end of reheating and $\rho_{e}$ is the energy density at the end of inflation. The number of degrees of freedom of species during reheating is denoted by $g_{re}$ and $g_{s,re}$ is the entropy number of degrees of freedom of species after reheating. The number of $e$-folds during radiation domination is given by
\beq
\label{NRD}
N_{rd}= -\frac{3(1+\omega_{re})}{4}N_{re}+\frac{1}{4} \ln[\frac{30}{g_{re} \pi^2}] +\frac{1}{3} \ln[\frac{11 g_{sre}}{43}]+\ln[\frac{a_{eq}\, \rho_e^{1/4}}{a_0\,T_0}].
\eeq
A final quantity of physical relevance is the thermalization temperature at the end of the reheating phase
\beq
\label{TRE}
T_{re}=\left( \frac{30\, \rho_e}{\pi^2 g_{re}} \right)^{1/4}\, e^{-\frac{3}{4}(1+\omega_{re})N_{re}}.
\eeq
\noindent 
Note that the condition $N_{re}\geq 0$ is equivalent to $T_{re}\leq T_{ins}$ where $T_{ins}=\left( \frac{30\, \rho_e}{\pi^2 g_{re}} \right)^{1/4}$ is the maximum allowed temperature corresponding to instantaneous reheating.
The last two equations define quantities given in terms of the number of $e$-folds during reheating thus, it is convenient to concentrate on $N_{re}$ given by Eq.~\eqref{NRE} which can also be written in a more convenient form for our purposes as follows
\beq
\label{NREfinal}
N_{re}=\left(1-3\, \omega_{re}\right)^{-1}\left(\ln\left[\frac{V_k}{V_e}re^{-4N_{ke}}
\right]+\ln\left[\frac{\pi^4A_s\, g_{re}}{270} \left(\frac{43}{11\,g_{s,re}}\right)^{4/3} \left(\frac{a_0 T_0}{k}\right)^{4}  \right]\right),
\eeq
where $V_k$ is the value of the potential when mode $k$ exits the Hubble radius during inflation and $V_e$ is the value of the potential at the end of inflation, $\rho_e=\frac{3}{2}V_e=\frac{9}{2}\frac{V_e}{V_k}H_k^2 M_{pl}^{2}=M_{pl}^{4}\frac{9\pi^2 A_s}{4}\frac{V_e}{V_k}r$ has been used. Only the first term in the rhs of Eq.~\eqref{NREfinal}
contains all the $n_s$ and $r$ dependence once the parameters of the model under consideration get eliminated in favor of $n_s$ and $r$. Numerical values are given as follows: $g_{s,re}=g_{re}=106.75$, $A_s=2.1\times 10^{-9}$, $T_0=2.725K=9.62\times 10^{-32}$, $k_p=0.05/Mpc=1.31\times 10^{-58}$, the last two quantities are also given by their dimensionless values in Planck units. We could obtain explicit formulas for $N_{re}$ for each of the models studied below but these are very cumbersome, instead we will find expressions for $V_k, V_e$ and $N_{ke}$ in terms of $n_s$ and $r$ and numerically find the relevant bounds for $n_s$ and $r$ by imposing conditions on $N_{re}$. 

For potentials where $V\sim \phi^n$ around the origin, an inflaton oscillating with frequency~$\sim a^{3(n-2)(n+2)}$ has an EoS given by \cite{Turner:1983he}
\begin{equation}
\omega = \frac{n-2}{n+2}\,.
\label{EoS}
\end{equation}
In what follows we illustrate our procedure by imposing the very general condition $N_{re}\geq 0$ (see Fig.~\ref{Nre}) to constrain $n_s$ and $r$ to models which are quadratic around the minimum thus, the EoS is given by $\omega=0$ in all the cases, and from there extract new bounds for cosmological quantities of interest. 
\section {\bf A class of $\alpha$-attractor models of inflation}\label{MOD}

The class of $\alpha$-attractor models we are interested in is given by the potential \cite{German:2021rin}
\begin{equation}
V=V_0\left(1-\sech^{p}(\lambda\frac{\phi}{M_{pl}})\right),
\label{potsech} 
\end{equation}
which is a different generalization from the basic potential given by the $p=2$ case \cite{Kallosh:2013yoa}
\begin{equation}
V=V_0\left(1-\sech^{2}(\lambda\frac{\phi}{M_{pl}})\right)=V_0\tanh^2(\lambda\frac{\phi}{M_{pl}})\,.
\label{potanh} 
\end{equation}
The generalization \eqref{potsech} follows \cite{Kallosh:2013yoa} in simplicity but adds the new requirement of having a 
positive definite function which gives viable potentials for any reasonable $p$ including odd and fractional values \cite{German:2021rin}. The potential \eqref{potsech} has also the interesting feature of being quadratic at the minimum for any $p$; around its minimum the potential \eqref{potsech} behaves like
\begin{equation}
V/V_0 =\frac{1}{2}p (\lambda \frac{\phi}{M_{pl}})^2-\frac{1}{24}p(2+3p)(\lambda \frac{\phi}{M_{pl}})^4+\cdot\cdot\cdot.
\label{potorigins}
\end{equation}
Solving the equation for the amplitude of scalar perturbations $A_s$ at $\phi_k$
\begin{equation}
A_s(k) =\frac{1}{24\pi ^{2}} \frac{V_k}{\epsilon_k\, M_{pl}^4}\,,
\label{Amp} 
\end{equation}
gives ($Se\equiv \sech(\lambda \frac{\phi_k}{M_{pl}})$)
\begin{equation}
Se=\left(1 - \frac{3}{2}\frac{M_{pl}^4}{V_0}A_s \pi^2 r\right)^{1/p}.
\label{Se} 
\end{equation}
which, however, involves the scale $V_0$. In terms of $Se$ the slow-roll parameters given by Eq.~\eqref{SR}, at $\phi=\phi_k$, are
\begin{eqnarray}
\epsilon_{_k} &=& \frac{p^2\,Se^{2p}(1-Se^2)}{2(1-Se^p)^2}\lambda^2, \label{Seps} \\
\eta_{_k} &=& \frac{p\,Se^p(Se^2-p(1-Se^2))}{1-Se^p}\lambda^2, \label{Seta} \\
\xi_{2_k} &=& \frac{p^2\,Se^{2p}(1-Se^2)(p^2-(1+p)(2+p)Se^2)}{(1-Se^p)^2}\lambda^4.\label{Segara}
\end{eqnarray}

From the equation $16\epsilon_k=r$ we get
\begin{equation}
\lambda=\left(\frac{r\left(1-Se^p\right)^2}{8p^2Se^{2p}(1-Se^2)}\right)^{1/2},
\label{lambda} 
\end{equation}
where $Se$ is given by Eq.~\eqref{Se} above. Thus, the value of the inflaton at horizon crossing is
\begin{equation}
\phi_k=\frac{M_{pl}}{\lambda}\arcsech \left(Se\right).
\label{fik} 
\end{equation}
It is easy to find that the running is now given by 
\begin{equation}
n_{sk} =\frac{3}{32}r^2 - \frac{1}{2} \delta_{n_s} r-\frac{1}{4}r\left(p^2-(1+p)(2+p)Se^2\right)\left(\frac{r(1-Se^p)^2}{8p^2Se^{2p}(1-Se^2)}\right).
\label{nskp} 
\end{equation}
To calculate quantities related with the radiation and reheating epochs as given by Eqs.~\eqref{NRD}, \eqref{TRE} and \eqref{NREfinal} we need expressions for $V_0$, $\phi_e$ and $N_{ke}$. Unfortunately it is not possible to solve for $V_0$ for a general $p$ thus, in what follows, we discuss a few particular cases originally considered in \cite{German:2021rin}. From the condition $N_{re}\geq0$ follows bounds for $n_s$ and $r$ (see Fig.~\ref{Nre}) and from there one can get bounds for other cosmological quantities.  The resulting bounds are given in the Table~\ref{bounds124}. Also, Figs.~\ref{Nre124} and \ref{nsk124} show $N_{re}$, $N_{ke}$, $N_{rd}$, $\log_{10} T_{re}$ and $n_{sk}$, as functions of $r$ for fixed values of $n_s$ within the constrained ranges for the three cases ($p=1 ,2, 4$) discussed in the text. Bounds for the case $n_s=0.9649$ corresponding to the mean value reported by the Planck collaboration \cite{Akrami:2018odb} are given in the Table \ref{0.9649}. In this case, we can further constrain the $p=1$ and $p=2$ models with the $p=4$ already ruled out although for smaller values of $n_s$ than 0.9649 the $p=4$  still defines a viable model.
\subsection {\bf The $p=1$ model}\label{p1}

Parameters such as $\lambda$, $V_0$ and most other expressions are specific to the particular example being solved and should not be used across subsections. For $p=1$ the potential \eqref{potsech} is given by \cite{Pal:2009sd}, \cite{Pal:2010eb}
\beq
\label{pot1}
V= V_0 \left(1- \sech\left(\lambda \frac{\phi}{M_{pl}}\right)\right).
\eeq
From Eq.~(\ref{Se})  we get
\beq
\label{Se1}
Se= 1 - \frac{3}{2}\frac{M_{pl}^4}{V_0}A_s \pi^2 r \,,
\eeq
and from Eq.~\eqref{lambda} follows that
\beq
\label{lambda1}
\lambda =  \frac{\sqrt{A_s}}{\sqrt{\frac{2}{3}\left(2\frac{V_0}{M_{pl}^4}-3A_s\pi^2 r \right)^2\left(4\frac{V_0}{M_{pl}^4}-3A_s\pi^2 r   \right)}}\frac{V_0}{M_{pl}^4}\pi\, r\,.
\eeq
The parameter $V_0$ follows from Eq.~\eqref{Ins} or $\delta_{n_s}+2\eta-\frac{3}{8}r=0$
\beq
\label{V01}
V_0 = \frac{3A_s \pi^2  r\left(24 \delta_{n_s}-r+\sqrt{17 r^2 +16 r \delta_{n_s} +64 \delta_{n_s}^2}\right)}{16(4 \delta_{n_s}-r)}M_{pl}^4\,,
\eeq
where, as before, $\delta_{n_s}\equiv 1-n_s$. Thus, from Eq.~\eqref{fik}
\beq
\label{fik1}
\phi_k =\frac{M_{pl}}{\lambda}\arcsech\left( \frac{3r-8 \delta_{n_s}+\sqrt{17 r^2 +16 r \delta_{n_s} +64 \delta_{n_s}^2}}{16\delta_{n_s}+2r}\right),
\eeq
while the end of inflation is given by the solution to the condition $\epsilon=1$
\beq
\label{fie1}
\phi_e =\frac{M_{pl}}{\lambda}\arcsech\left(-\frac{1}{3}+\frac{6-\lambda^2-R_1^{2/3}}{3\lambda R_1^{1/3}}\right),
\eeq
where $R_1\equiv -36\lambda+\lambda^3+3\sqrt{6}\sqrt{4+22\lambda^2-\lambda^4}$ with $\lambda$ given by Eq.~\eqref{lambda1}. The number of $e$-folds during inflation is
\beq
\label{Nke1}
N_{ke} =\frac{1}{\lambda^2}\left(\cosh\left(\lambda\frac{\phi_k}{M_{pl}}\right)-\cosh\left(\lambda\frac{\phi_e}{M_{pl}}\right)-2 \ln\left[\cosh\left(\lambda\frac{\phi_k}{2M_{pl}}\right)\right]+2 \ln\left[\cosh\left(\lambda\frac{\phi_e}{2M_{pl}}\right)\right]
\right).
\eeq
Finally, from Eqs.~\eqref{nskp}, \eqref{Se1} and \eqref{V01} the running $n_{sk}$ can be written as follows
\begin{equation}
n_{sk} = \frac{1}{256}\left(51 r^2 + 80 r  \delta_{n_s} - 64 \delta_{n_s}^2-(13 r + 8  \delta_{n_s})\sqrt{17 r^2 +16 r \delta_{n_s} +64 \delta_{n_s}^2}\right).
\label{nsksolmu}
\end{equation}
\begin{figure}[t!]
\begin{center}
\includegraphics[trim = 0mm  0mm 1mm 1mm, clip, width=10.cm, height=7.cm]{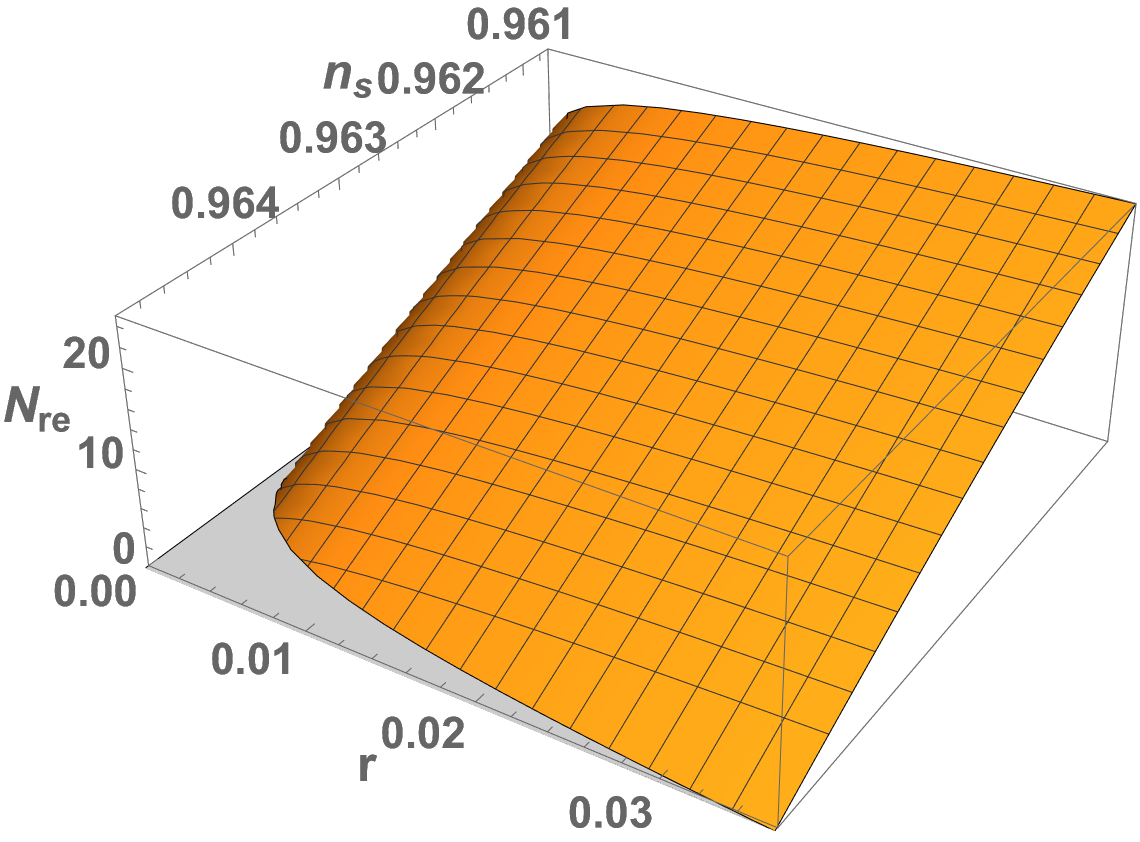}
\end{center}
\caption{Plot of the number of $e$-folds during reheating as a function of $n_s$ and $r$ for the $p=2$ model and equation of state $\omega_{re}=0$. This EoS corresponds to the quadratic potentials around their minimum which follow from Eq.~\eqref{potsech}. The original range for the scalar spectral index is $0.9607 < n_s < 0.9691$ \cite{Akrami:2018odb} and $r < 0.036$ for the tensor-to-scalar ratio \cite{BICEPKeck:2021gln}. The condition $N_{re}\geq 0$ restricts the values of $n_s$ and $r$ further to the ranges $0.9607<n_s<0.9650$ and $8.9 \times 10^{-11} < r < 0.036$ respectively. From these bounds we obtain in turn new bounds for the number of $e$-folds during inflation $N_{ke}$, radiation $N_{rd}$, the reheating temperature $T_{re}$ and for the running index $n_{sk}$ as given in the Table~\ref{bounds124}. The reader will notice that the number of $e$-folds during reheating depends strongly on the spectral index $n_s$ but not so much on the tensor-to-scalar ratio $r$. Even if we knew precisely the value of $r$, a spread of values in $n_s$ as shown in the figure would lead to an uncertainty of approximately 20 $e$-folds during reheating. Something equivalent can be said in relation to the other cosmological quantities quoted above.
}
\label{Nre}
\end{figure}
Having obtained the parameters $\lambda$ and $V_0$ in terms of $n_s$ and $r$ we can proceed to obtain the bounds for $n_s$ and $r$ by imposing the condition $N_{re}\geq 0$  (see Fig.~\ref{Nre}) and from there obtain bounds for other quantities of interest. The resulting bounds for the $p=1$ model are given by the second column of Table \ref{bounds124}. 
\subsection {\bf The $p=2$ model}\label{p2}

The $p=2$ model is given by the potential 
\beq
\label{pot2}
V= V_0 \left(1- \sech^2\left(\lambda \frac{\phi}{M_{pl}}\right)\right).
\eeq
We apply the procedure discussed in the previous subsection to obtain $V_0$ with the result
\beq
\label{V02}
V_0 = \frac{6A_s \pi^2  r \delta_{n_s}}{4\delta_{n_s}-r}M_{pl}^4\,.
\eeq
The parameter $\lambda$ is
\begin{equation}
\lambda= \sqrt{\frac{\delta_{n_s}\left(4\delta_{n_s}-r\right)}{8r}}\;.
\label{lambda2}
\end{equation}
Thus,
\beq
\label{fik2}
\phi_k =\sqrt{\frac{8r}{\delta_{n_s}\left(4\delta_{n_s}-r\right)}}M_{pl} \arcsech\left( \frac{1}{2}\sqrt{\frac{r}{\delta_{n_s}}}\right)\,.
\eeq
The end of inflation is given by the condition $\epsilon=1$ at $\phi_e$ where $\phi_e$ is given by
\beq
\label{fie2}
\phi_e =\sqrt{\frac{8r}{\delta_{n_s}\left(4\delta_{n_s}-r\right)}}M_{pl} \arcsech\left( \frac{\sqrt{2}\,r^{1/4}\left(\sqrt{r+\delta_{n_s}\left(4\delta_{n_s}-r\right)}-\sqrt{r}\right)^{1/2}}{\sqrt{\delta_{n_s}\left(4\delta_{n_s}-r\right)}}\right)\,.
\eeq
The number of $e$-folds during inflation is 
\beq
\label{Nke2}
N_{ke} =\frac{8\delta_{n_s}-r-\sqrt{r^2+r\delta_{n_s}(4\delta_{n_s}-r)}}{\delta_{n_s}(4\delta_{n_s}-r)}.
\eeq
The running is simply given by 
\begin{equation}
n_{sk} = - \frac{\delta_{n_s}^2}{2}\;,
\label{nsk2}
\end{equation}
and does not depend on $r$. Results for the $p=2$ model are given by the third column of Table \ref{bounds124}.
 \begin{center}
\scriptsize
\begin{table*}[htbp!]
\resizebox{\textwidth}{!}{\begin{tabular}{cccc}
\small
$Characteristic$ & $p=1$ & $p=2$ & $p=4$\\ \hline\\[0.1mm]
$n_s$   &     $0.9674 > n_{s} > 0.9607 $ & $0.9650 > n_{s} > 0.9607 $ & $0.9635 > n_{s} > 0.9607 $\\[2mm] 
$r$   &   $4.27\times 10^{-4}< r <0.036$ & $8.90\times 10^{-11}< r <0.036$ & $8.94\times 10^{-11}< r <0.036$\\[2mm]
$n_{sk}$  & $\left(-5.8\times 10^{-4}, -8.4\times 10^{-4}\right)$ & $\left(-6.1\times 10^{-4}, -7.7\times 10^{-4}\right)$ & $\left(-6.2\times 10^{-4}, -7.2\times 10^{-4}\right)$\\[2mm]
$\lambda$   &   $ 2.68> \lambda > 0.167$ & $ 2946 > \lambda >0.1123  $ & $ 1470 > \lambda > 0.0774 $\\[2mm]
$N_{ke}$   &   $56.4 > N_{ke} > 46.7$ & $56.7> N_{ke} >50.4 $ & $56.7> N_{ke} >52.8 $\\[2mm]
$N_{re}$  &    $0< N_{re}< 38.8$ & $0< N_{re}< 25.0$ & $0< N_{re}< 15.6$\\[2mm]
$N_{rd}$  &   $57.6 > N_{rd} > 28.6$ & $57.6 > N_{rd} > 38.9$ & $57.6 > N_{rd} > 45.9$\\[2mm]
$T_{re} (GeV)$  &  $\left(2.8\times 10^{15}, 7.1\times 10^{2}\right)$ &  $\left(2.7\times 10^{15}, 2.1\times 10^{7}\right)$ &  $\left(2.7\times 10^{15}, 2.2\times 10^{10}\right)$ \\[2mm]
\end{tabular}}
\caption{\label{bounds124} The scalar spectral index $n_s$ and the tensor-to-scalar ratio $r$ have been constrained by the condition $N_{re}\geq 0$ and these are then used to bound all the other quantities appearing in the table. We collect bounds for various cosmological quantities of interest such as the running index $n_{sk}$, the number of $e$-folds during inflation $N_{ke}$, reheating $N_{re}$ and radiation $N_{rd}$ together with the reheating temperature $T_{re}$. The way inequalities are written suggests relations among the different quantities. Thus, the lower bound for $\lambda$ implies an upper bound for the tensor-to-scalar ratio $r$ which in turn determines the upper bound for the number of $e$-folds during reheating.
}
\end{table*}
\end{center}
\subsection {\bf The $p=4$ model}\label{p4}

As in the previous subsections one can get 
\beq
\label{V04}
V_0 = \frac{3A_s \pi^2  r R_2}{32\left(8\delta_{n_s}-3r\right)\left(4\delta_{n_s}-r\right)}M_{pl}^4\,,
\eeq
from where it follows that
\begin{equation}
\lambda=\frac{\sqrt{2r}\left(8\delta_{n_s}-3r\right)\left(4\delta_{n_s}-r\right)}{R_3\sqrt{1-\sqrt{R_3/R_2}}}\;,
\label{lambda4}
\end{equation}
where $R_2\equiv 512\delta_{n_s}^2-192r\delta_{n_s}+9r^2-r\sqrt{r(256\delta_{n_s}-15r)}$ and  $R_3\equiv 128r\delta_{n_s}-39r^2-r\sqrt{r(256\delta_{n_s}-15r)}$.
At horizon crossing the inflaton is 
\beq
\label{fik4}
\phi_k =\frac{M_{pl}}{\lambda}\arcsech\left(1-\frac{3M_{pl}^4}{2V_0}A_s \pi^2 r
\right)^{1/4}\,,
\eeq
while the end of inflation is given as usual by the saturation of the condition $\epsilon=1$
\beq
\resizebox{1.00\textwidth}{!}{$
\label{fie4}
\phi_e =\frac{M_{pl}}{\lambda}\arcsech\left(\frac{1}{4\sqrt{6}\lambda}\left(-3+\sqrt{9-192\lambda^2+3R_4}+\sqrt{18-384\lambda^2-3R_4-\frac{18\sqrt{3}(1-32\lambda^2-512\lambda^4)}{\sqrt{3-64\lambda^2+R_4}}}\right)^{1/2}
\right)\,,$}
\eeq
$R_4\equiv \frac{32\lambda^2\left(2-48\lambda^2+2^{1/3}\left(-2+3\lambda\left(33\lambda+\sqrt{-12+321\lambda^2+6144\lambda^4}\right)\right)^{2/3}\right)}{\left(-1+\frac{3}{2}\lambda\left(33\lambda+\sqrt{-12+321\lambda^2+6144\lambda^4}\right)\right)^{1/3}}$, with $\lambda$ now given by Eq.~\eqref{lambda4} above. The number of $e$-folds during inflation is
\beq
\label{Nke4}
N_{ke} =\frac{1}{128\lambda^2}\left(12\cosh\left(2\lambda\frac{\phi_k}{M_{pl}} \right)-12\cosh\left(2\lambda\frac{\phi_e}{M_{pl}} \right)
+\cosh\left(4\lambda\frac{\phi_k}{M_{pl}} \right)-\cosh\left(4\lambda\frac{\phi_e}{M_{pl}} \right)\right).
\eeq
Finally,  Eq.~\eqref{nskp}  implies  
\begin{equation}
n_{sk} =\frac{3}{32}r^2 - \frac{1}{2} \delta_{n_s} r -  \frac{r^2 R_5\left(15R_5-7(1+\sqrt{1-R_5})\right)}{256(1-R_5)^2}\;,
\label{nsk4}
\end{equation}
where $R_5\equiv 3A_s \pi^2 r M^4_{pl}/2V_0$.
The resulting bounds for the $p=4$ model are given in the last column of Table \ref{bounds124}. 
\begin{figure*}
\begin{center}$
\begin{array}{cccc}
\small
\includegraphics[width=1.5in]{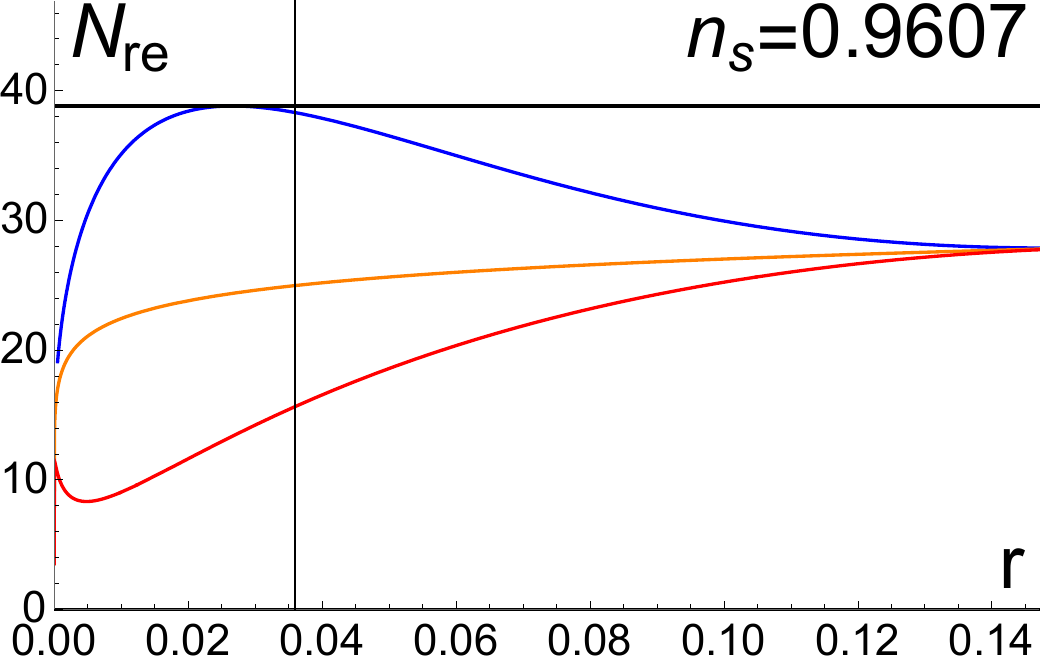}&
\includegraphics[width=1.5in]{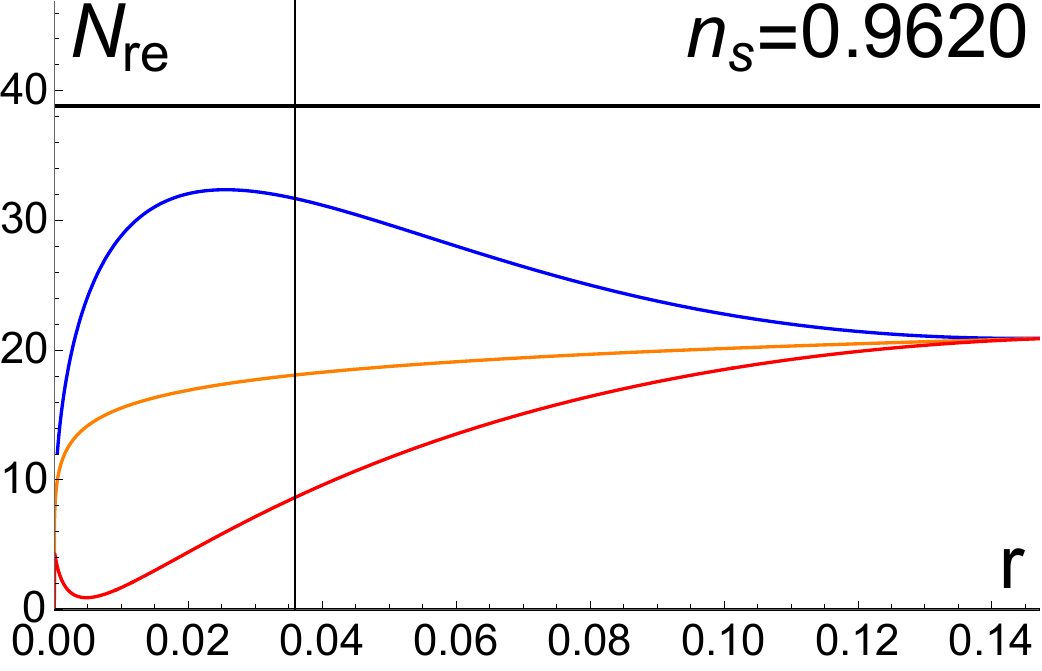}&
\includegraphics[width=1.5in]{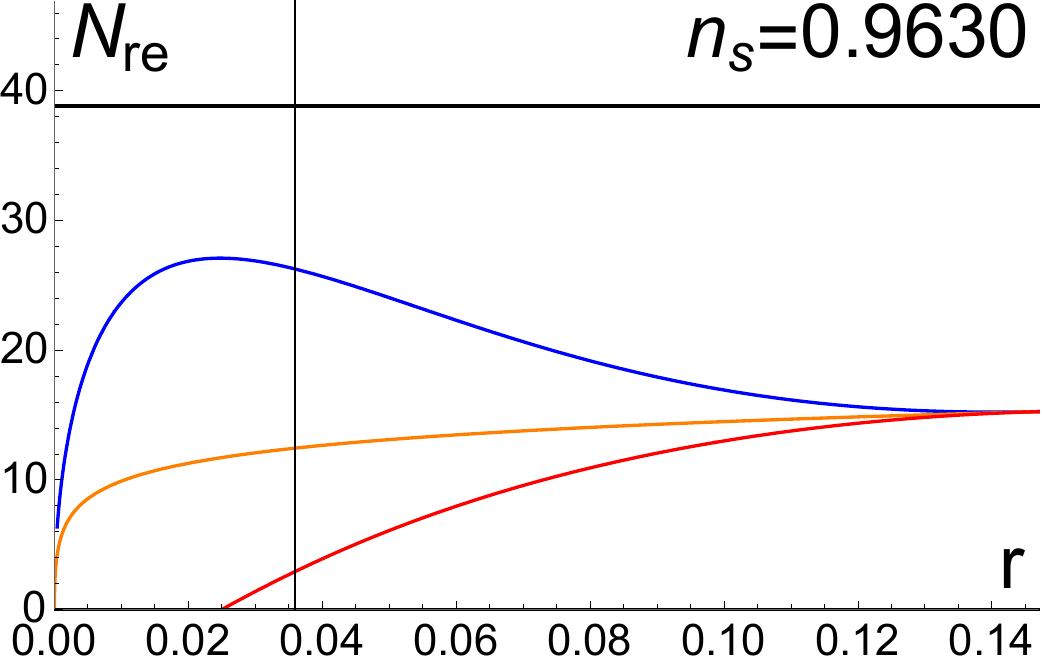}&
\includegraphics[width=1.5in]{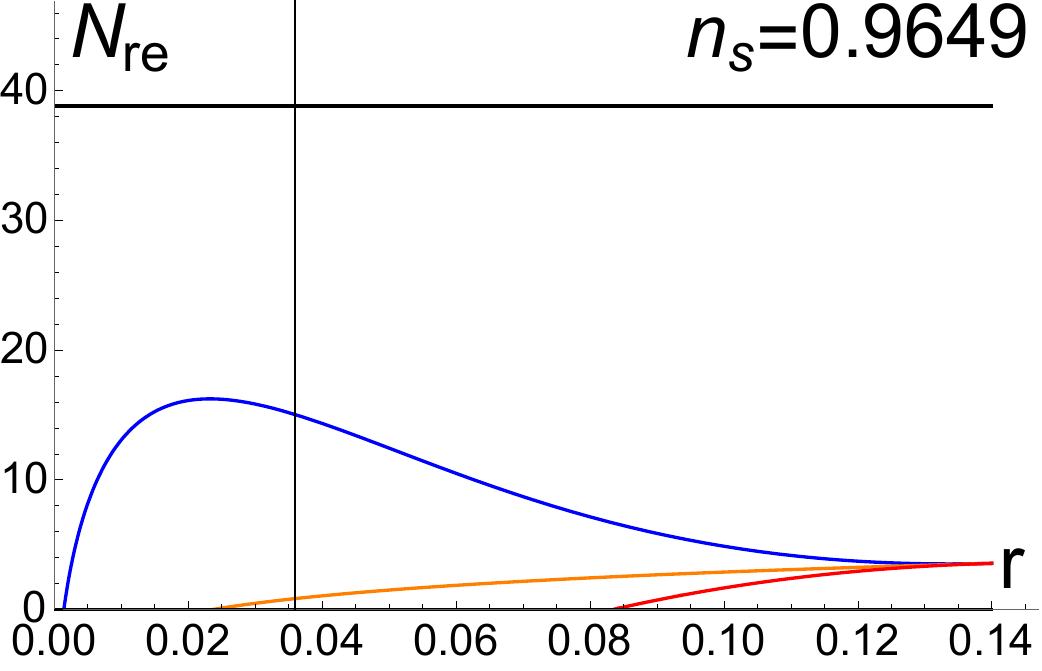}\\
\includegraphics[width=1.5in]{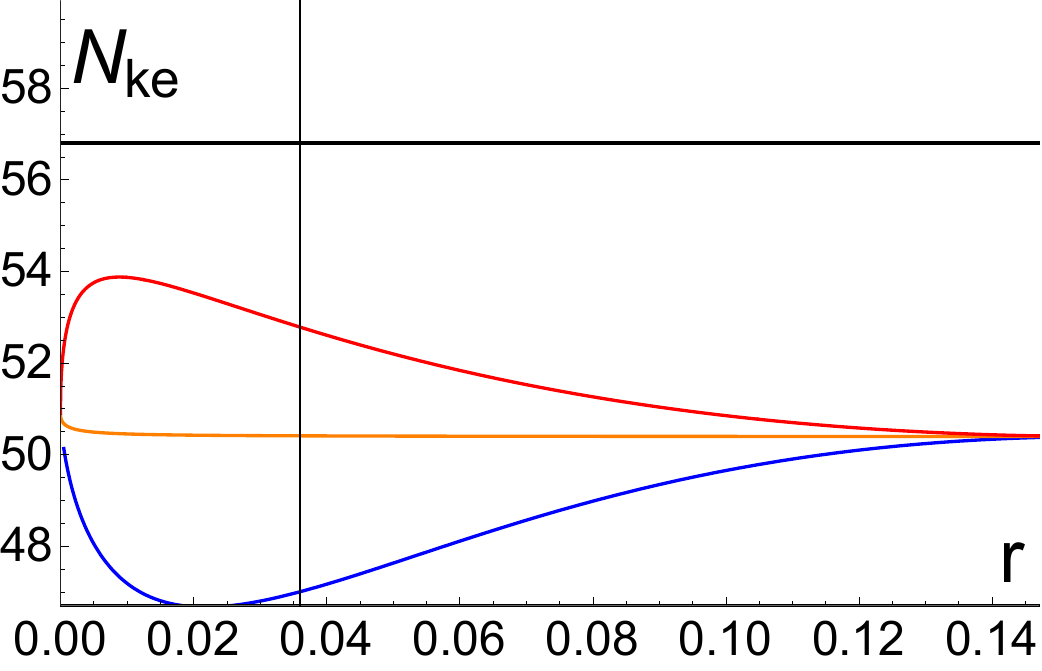}&
\includegraphics[width=1.5in]{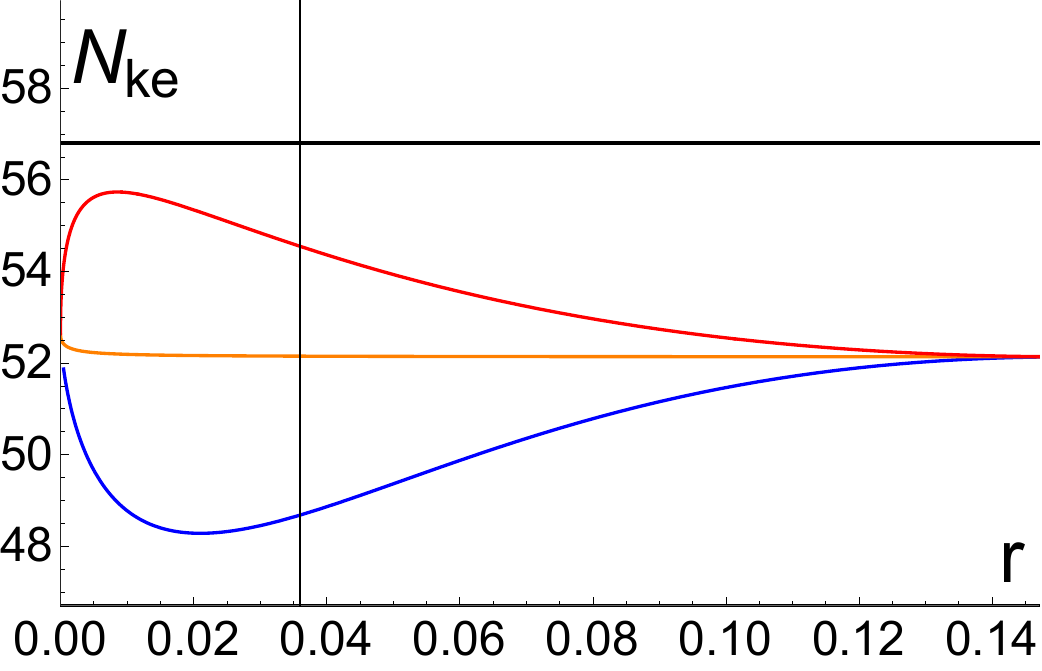}&
\includegraphics[width=1.5in]{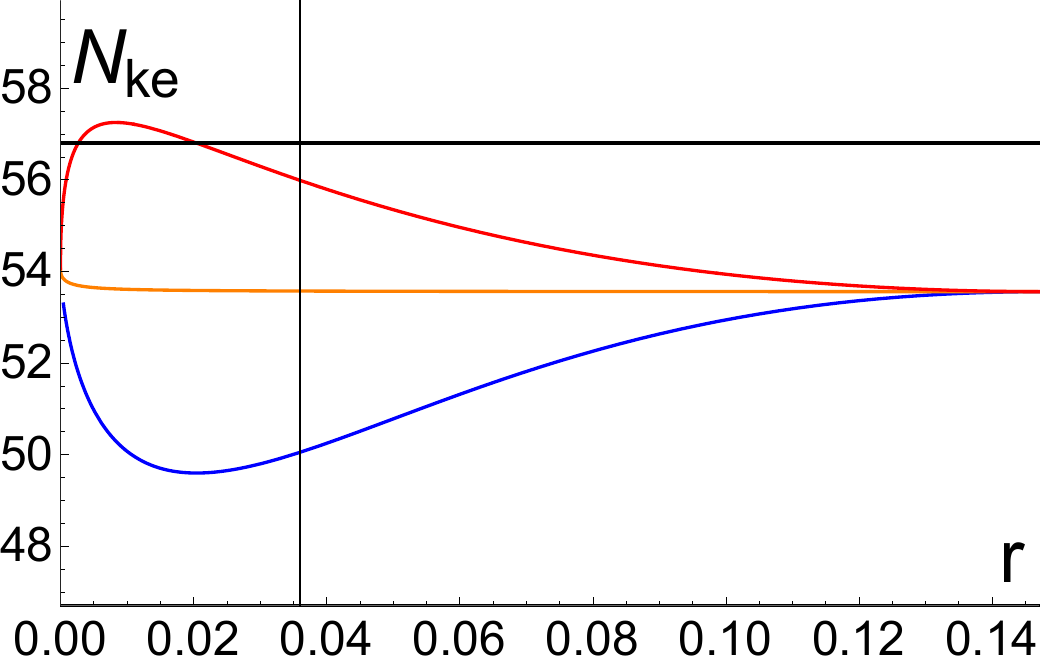}&
\includegraphics[width=1.5in]{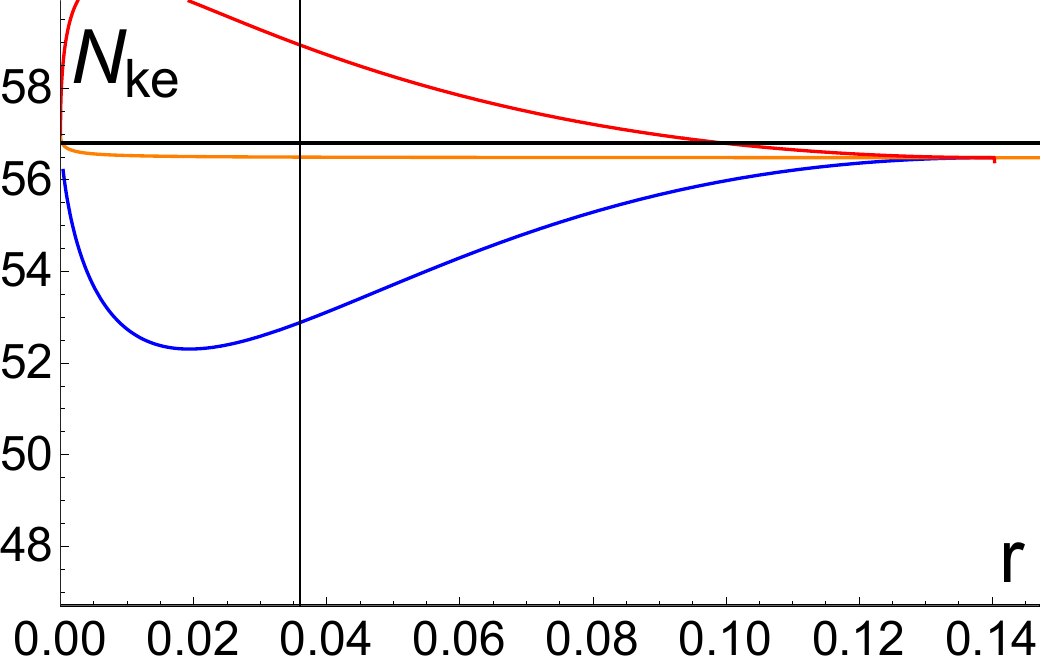}\\
\includegraphics[width=1.5in]{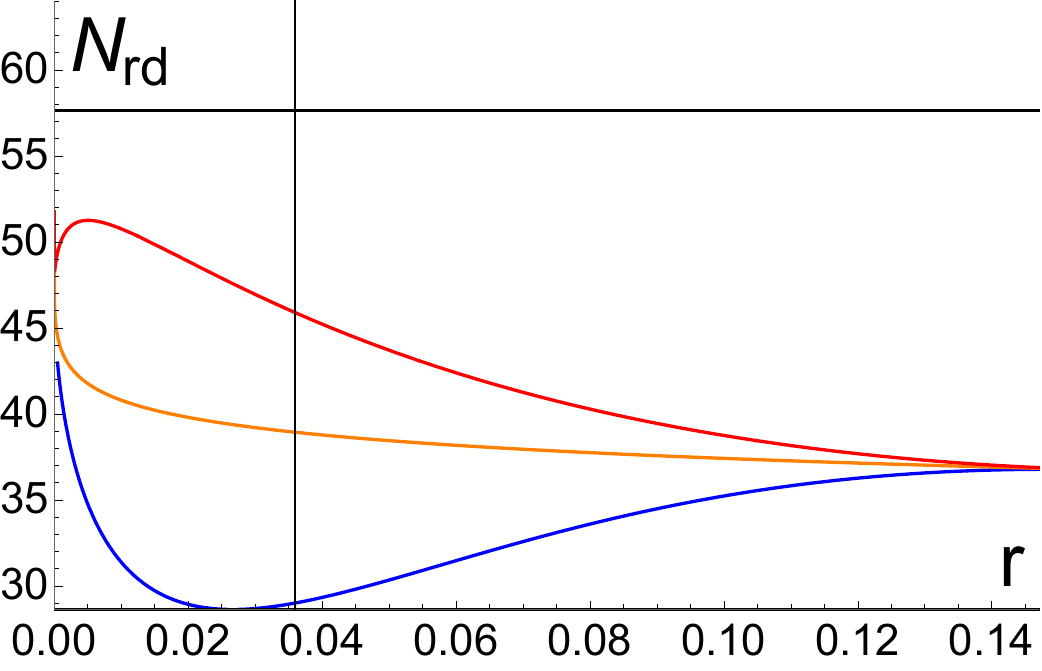}&
\includegraphics[width=1.5in]{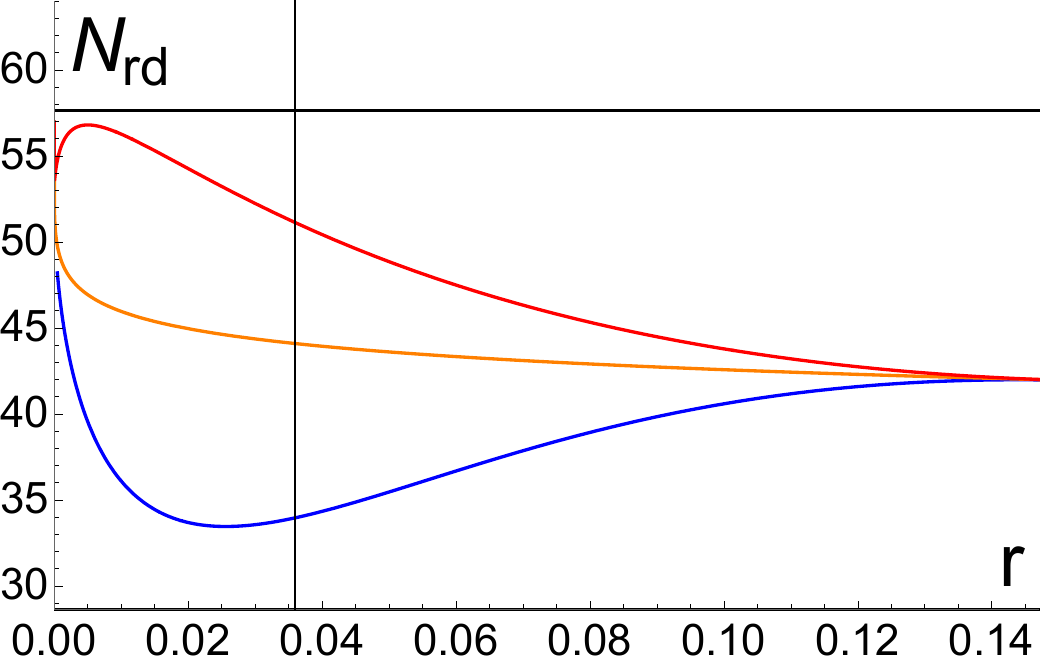}&
\includegraphics[width=1.5in]{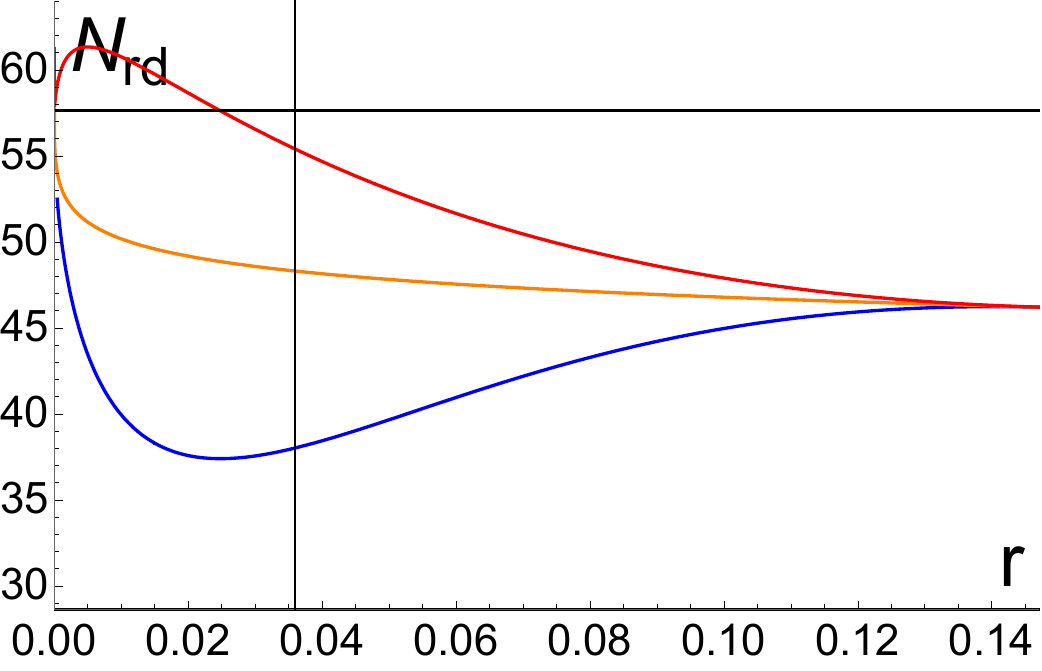}&
\includegraphics[width=1.5in]{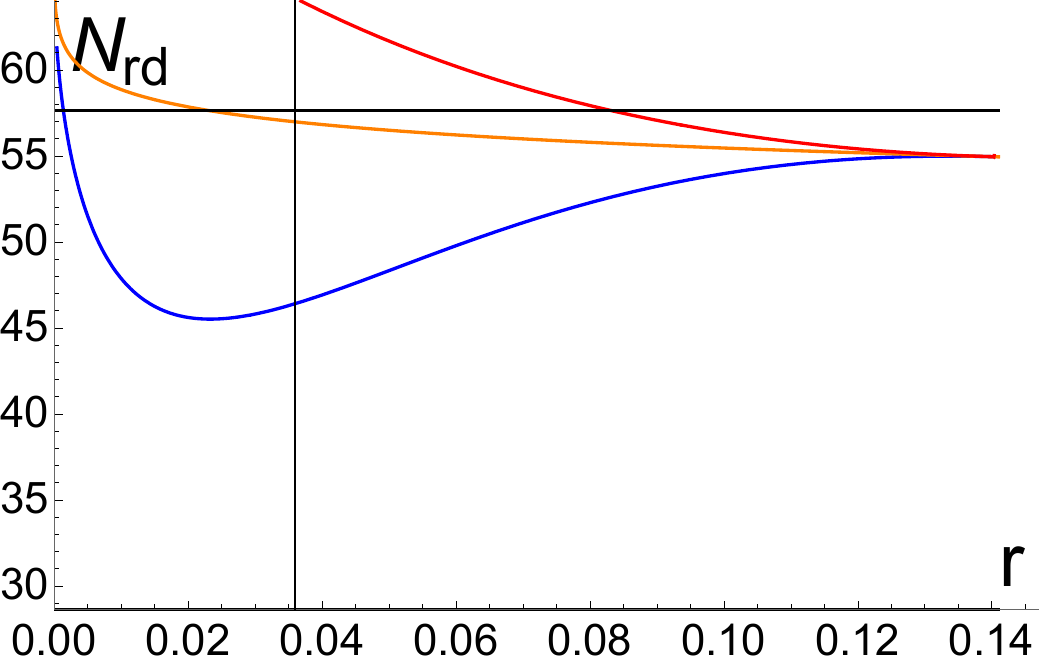}\\
\includegraphics[width=1.5in]{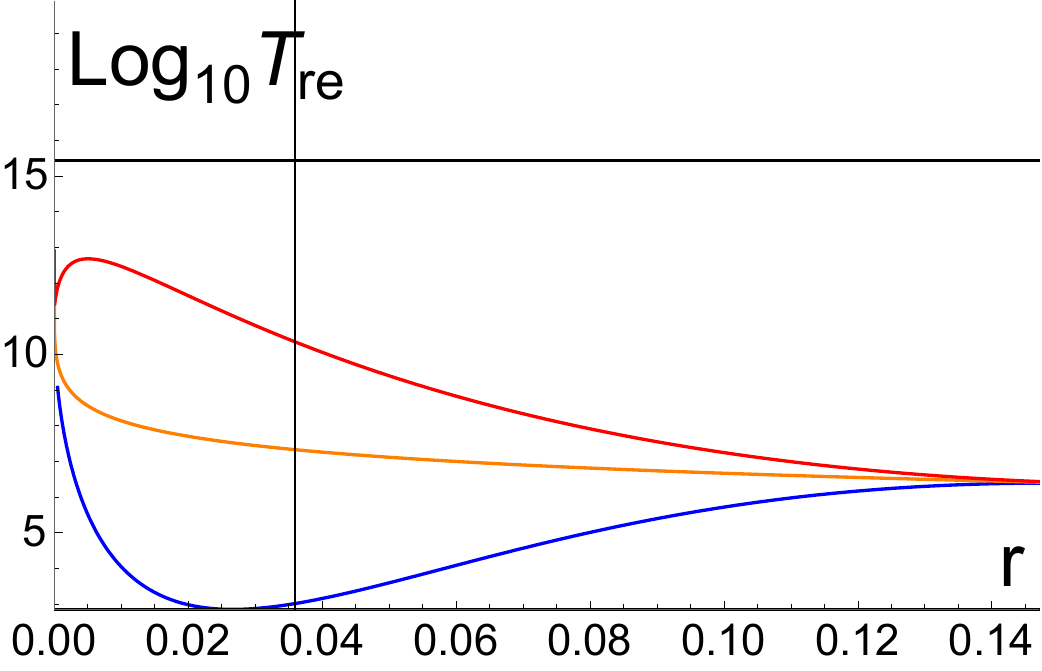}&
\includegraphics[width=1.5in]{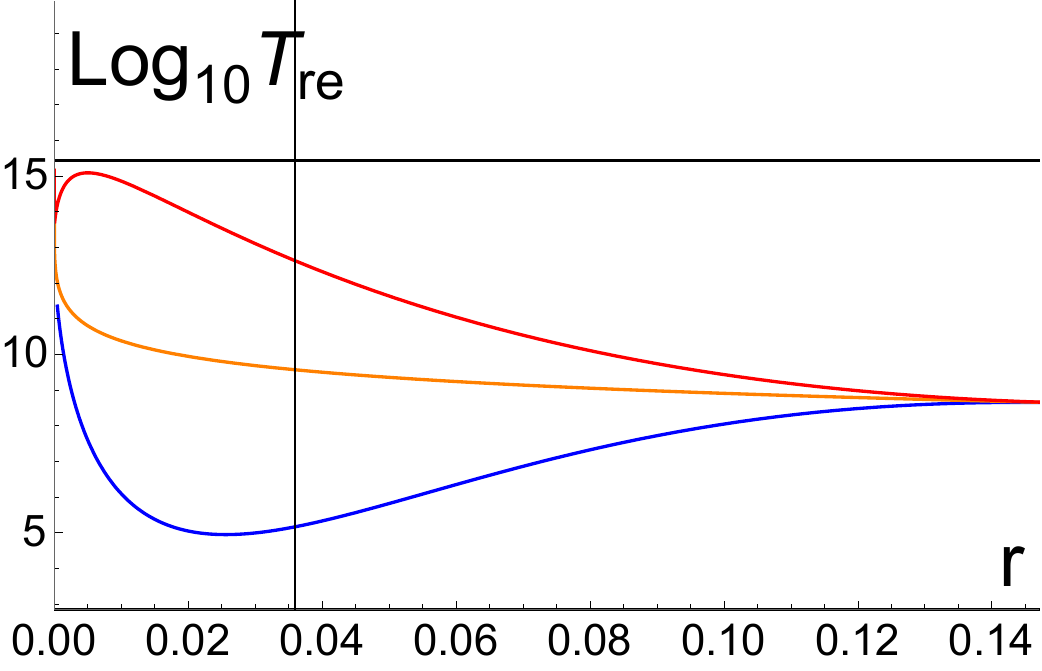}&
\includegraphics[width=1.5in]{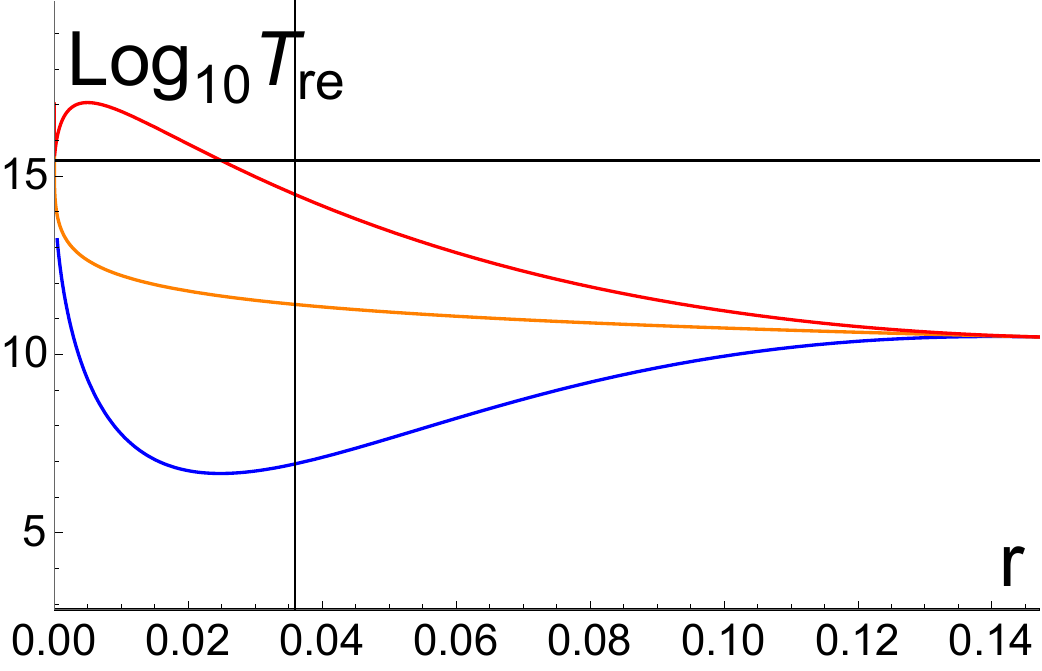}&
\includegraphics[width=1.5in]{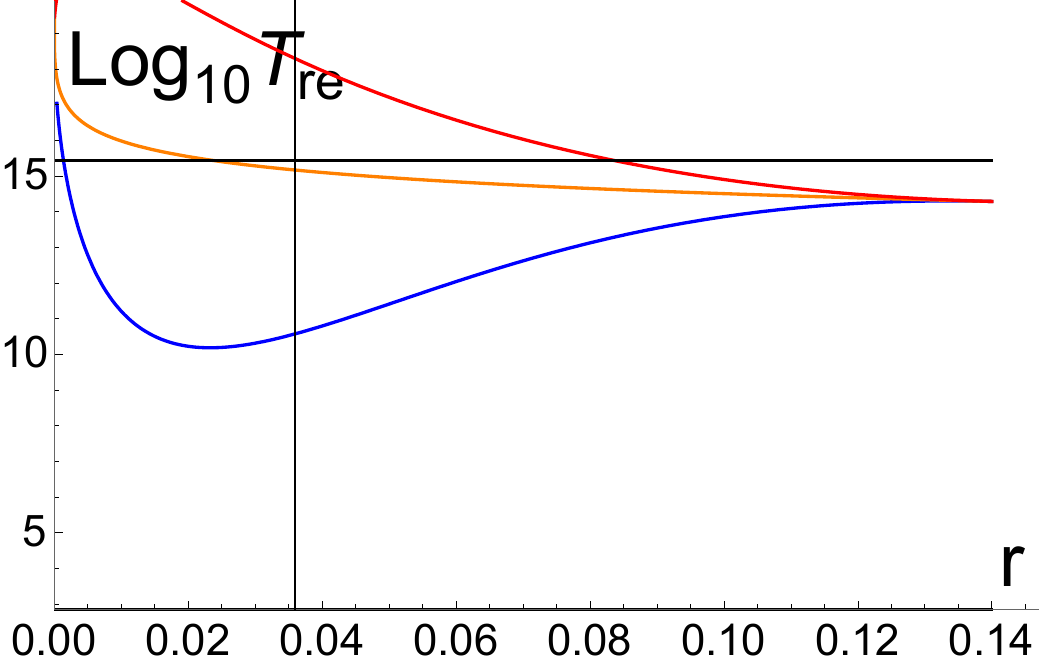}
\end{array}$
\end{center}
\caption{Plot of the number of $e$-folds during reheating $N_{re}$, inflation $N_{ke}$ and radiation  $N_{rd}$ together with the $log_{10}$ of the reheating temperature $T_{re}$ as functions of the tensor-to-scalar ratio $r$ for various values of the spectral index $n_s$. In each panel the horizontal lines signal the extreme bounds for the plotted quantities while the vertical line corresponds to the recently obtained $r=0.036$ bound \cite{BICEPKeck:2021gln}.
We can consider the array of figures as a $4\times 4$ matrix (denoted in what follows by $M_{ij}$, $i, j = 1,2,3,4$) thus, columns correspond to the same value of $n_s$ specified at the top, increasing along the row. In the first row $M_{1j}$ the blue (top) curve corresponds to the $p=1$ case with the orange (middle) curve to $p=2$ and red to $p=4$.  This is inverted in rows $M_{2j}$, $M_{3j}$ and $M_{4j}$ with the blue (bottom) curve for the $p=1$ model. In the panel $M_{14}$, corresponding to $N_{re}$ for the mean value  $n_s=0.9649$ \cite{Akrami:2018odb}, we see that the  $p=4$ model is already ruled out with $r>0.08$ and the models $p=1$ and $p=2$ further restricted as shown in the Table~\ref{0.9649}. The model defined by $p=4$ is still viable for smaller values of $n_s$ as shown in the panels $M_{11}$, $M_{12}$ and narrowly in the panel $M_{13}$. The point where the curves join in the extreme right of the figures corresponds with the transition of the potentials \eqref{potsech} to the monomial $\phi^2$ as shown in \cite{German:2021rin}.}
\label{Nre124}
\end{figure*}
\begin{figure*}
\begin{center}$
\begin{array}{cccc}
\includegraphics[width=1.5in]{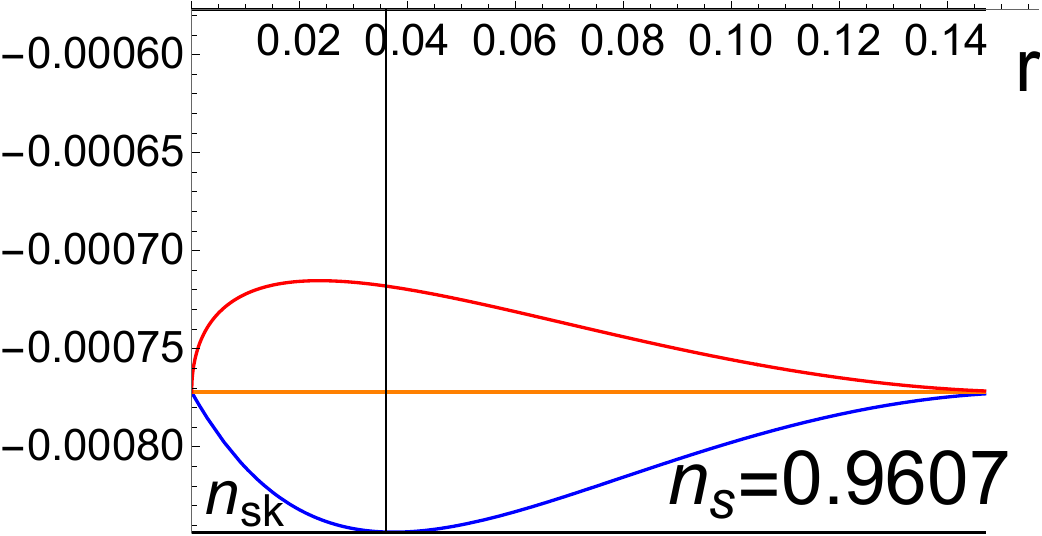}&
\includegraphics[width=1.5in]{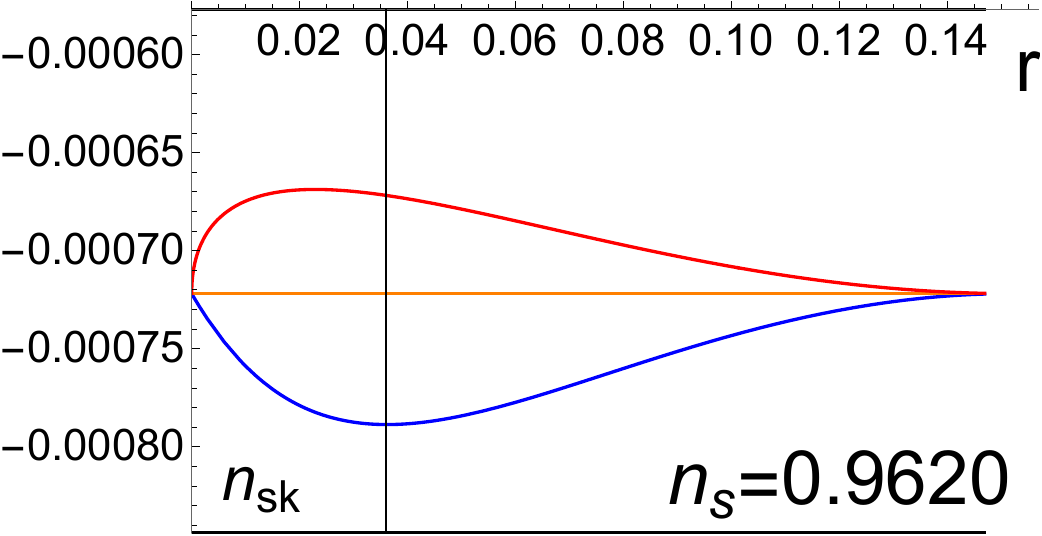}&
\includegraphics[width=1.5in]{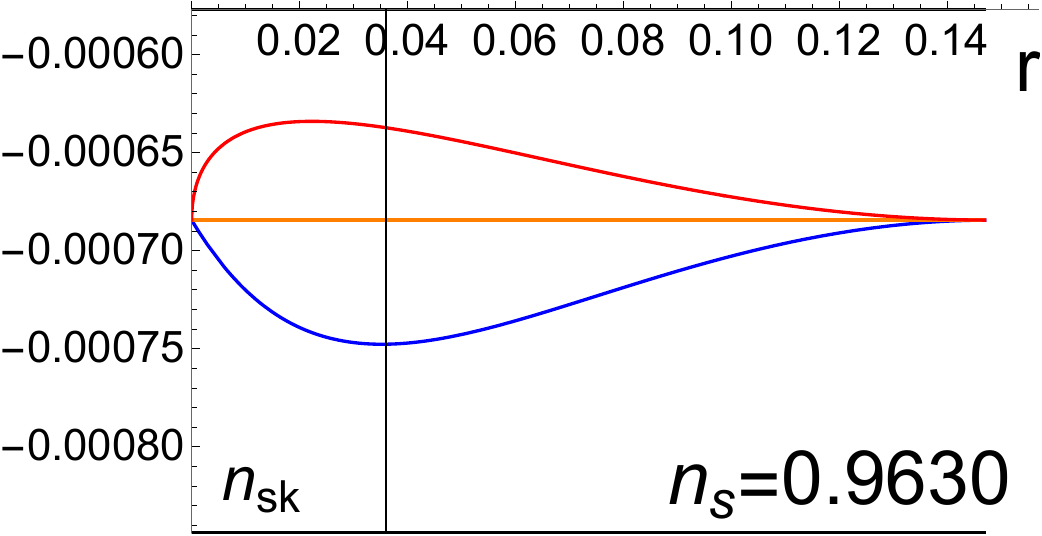}&
\includegraphics[width=1.5in]{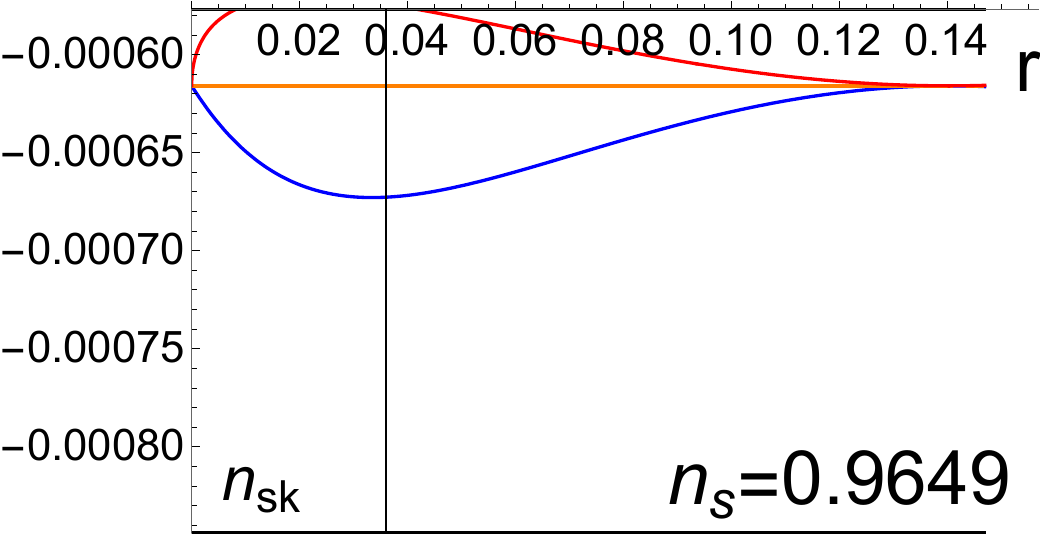}
\end{array}$
\end{center}
\caption{Plot of the running index $n_{sk}$ as a function of the tensor-to-scalar ratio $r$ for various values of the scalar spectral index $n_s$ as indicated. In each panel the horizontal lines signal the extreme bounds for $n_{sk}$ while the vertical line corresponds to the $r=0.036$ bound \cite{BICEPKeck:2021gln}.}
\label{nsk124}
\end{figure*}
 \begin{center}
\addtolength{\tabcolsep}{0 pt}
\begin{table*}[htbp!]
\begin{tabular}{ccc}
$Characteristic$ & $p=1$ & $p=2$ \\ \hline\\[0.1mm]
$n_s$   &     $0.9649$ & $0.9649$\\[2mm] 
$r$   &   $0.00144< r <0.036$ & $0.02370< r <0.036$\\[2mm]
$n_{sk}$  & $-6.22\times 10^{-4} > n_{sk} > -6.69\times 10^{-4}$ & $-6.16\times 10^{-4}$\\[2mm]
$\lambda$   &   $0.2565 < \lambda < 1.29$ & $0.1128 < \lambda < 0.1470 $\\[2mm]
$N_{ke}$   &   $55.3 > N_{ke} > 52.3$ & $56.5$\\[2mm]
$N_{re}$  &    $0< N_{re}< 16.2$ & $0< N_{re}< 0.84$\\[2mm]
$N_{rd}$  &   $57.4 > N_{rd} > 45.5$ & $57.6 > N_{rd} > 57.0$\\[2mm]
$T_{re} (GeV)$  &  $2.3\times 10^{15} > T_{re} > 1.5\times 10^{10}$ &  $2.7\times 10^{15}  > T_{re} > 1.5\times 10^{15}$\\[2mm]
\end{tabular}
\caption{\label{0.9649} In the panel $M_{14}$ of Fig.~\ref{Nre124}, corresponding to $N_{re}$ for the mean value $n_s=0.9649$ as given by the Planck 2018 collaboration article \cite{Akrami:2018odb},  we see that the case $p=4$ is already ruled out with $r>0.08$. The models defined by $p=1$ and $p=2$ are further restricted by the condition $N_{re}\geq0$. The corresponding bounds for $r$, the running index $n_{sk}$, the number of $e$-folds during inflation $N_{ke}$, reheating $N_{re}$ and radiation $N_{rd}$ together with the reheating temperature $T_{re}$ are given above.
}
\end{table*}
\end{center}
\section {\bf Consistency of the results}\label{CONSISTENCY}
A formula for the total number of $e$-folds from the time scales with wavenumber mode $k_p\equiv a_p H_p$ exit the horizon during inflation until that same scale re-enters the horizon has been given in \cite{German:2020iwg}. This formula can be written as
\begin{equation}
N_{kp}\equiv N_{ke}+N_{ep}=\ln[\frac{a_{p}H_k}{k_p}]\;,
\label{EQ}
\end{equation}
where $a_p$ is the scale factor at the pivot scale given by $a_p=3.6512\times 10^{-5}$ \cite{German:2020iwg}, $N_{ke}\equiv \ln\frac{a_e}{a_k}$ is the number of $e$-folds during inflation from $\phi_k$ up to the end of inflation at $\phi_e$  and $N_{ep}\equiv  \ln\frac{a_p}{a_e}$ is the {\it postinflationary} number of $e$-folds from the end of inflation at $a_e$ up to reentry where the pivot scale factor is $a_p$. Note that $N_{ep}=N_{re}+N_{rd}-2.1$ i.e., $N_{ep}$ is the number of e-folds during reheating plus the number of e-folds during radiation minus 2.1 e-folds because the universe has expanded 2.1 e-folds from the time scales the size of the pivot scale re-enter the horizon to the time of equality, that is $\ln\frac{a_{eq}}{a_p}=2.1$, \cite{German:2020iwg}.

Writing the Hubble function at $k$ as $H_k=M_{pl}\sqrt{\frac{\pi^2 A_s}{2} r}$,  Eq.~\eqref{EQ} can also be written as
\begin{equation}
N_{kp}=\frac{1}{2}\ln\left(\frac{M_{pl}^{2}\pi^2 a_p^2 A_s}{2 k_p^2}r\right)\;.
 \label{EQbound}
\end{equation}
As an example of the consistency of our results with Eq.~\eqref{EQbound} we calculate the bounds on $N_{kp}=N_{ke}+N_{re}+N_{rd}-2.1$ for the $p=2$  as given by the third column of the Table~\ref{bounds124}. In this case the addition of the lhs bounds is 56.7+0+57.6-2.1=112.2 and the addition of the rhs bounds 50.4+25.0+38.9-2.1=112.2. On the other hand  Eq.~\eqref{EQbound} at $r=0.036$ gives 112.2 in complete agreement with the calculated bounds as explained in the text. The same is true for the other models studied here. Minute differences of ${\cal O}(0.1)$ $e$-folds are due to the rounding up.

Another consistency check occurs for the bounds on the number of $e$-folds for each of the epochs considered  and displayed in the Table~\ref{bounds124} by comparing with results obtained from a diagrammatic approach studied in \cite{German:2020kdp}. From the third column of Table I of Ref.~\cite{German:2020kdp} we find that (in the notation used here) $56.2>N_{ke}>42.4$, $0<N_{re}<55.5$ and $58.3>N_{rd}>16.7$. One can easily check that all the results presented here fall, with uncertainties of a few tenths of an e-fold due to the choice of the minimum reheat temperature in \cite{German:2020kdp}, within these bounds. 
\section {\bf Conclusions}\label{CON}
We have proposed a procedure by which we can establish bounds for cosmological quantities of interest. The key point consist in eliminating the parameters introduced by the potential of the model in terms of the observables $n_s$ and $r$. In particular, we have concentrated on potentials quadratic in the inflaton around the minimum where reheating takes place and have imposed the very general condition $N_{re}\geq 0$ for the number of $e$-folds during reheating before entering the radiation era. 

We have used three inflationary models from a class of $\alpha$-attractor models to illustrate the suggested procedure as fully as possible. We present our results in Figs.~\ref{Nre124} and \ref{nsk124} and the tables \ref{bounds124} and \ref{0.9649}. We test the consistency of the results with formulae and procedures studied previously. It is clear that this approach can be used to discard models of inflation as discussed in Section~\ref{MOD} and the Table~\ref{0.9649}. An interesting result is that in the three models the resulting tensor-to-scalar ratio $r$ (and as a consequence the energy scale of inflation)  is bounded from below as follows from Eq.~\eqref{NREfinal}.
\acknowledgments
I would like to thank J.C. Hidalgo, F.X. Linares Cedeño, A. Montiel and J. Ferrara for useful conversations. We acknowledge financial support from UNAM-PAPIIT,  IN104119, {\it Estudios en gravitaci\'on y cosmolog\'ia}.

\end{document}